# Topoisomer differentiation of molecular knots by FT-ICR-MS: lessons from class II lasso peptides

AUTHOR NAMES

*Séverine Zirah*[*,1,§], *Carlos Afonso*[*,2], *Uwe Linne*[3], *Thomas A. Knappe*[3], *Mohamed A. Marahiel*[3], *Sylvie Rebuffat*[1] *and Jean-Claude Tabet*[2]

* these authors have contributed equally to the work

§ Corresponding author

AUTHOR ADDRESS

1. National Museum of Natural History, Communication Molecules and Adaptation of Micro-organisms, FRE 3206 CNRS - MNHN, CP 54, 57 rue Cuvier, F-75005 Paris, France

2. Parisian Institute of Molecular Chemistry, UMR 7201 CNRS - University Pierre & Marie Curie, 4 place Jussieu, F-75005 Paris, France.

3. Department of Chemistry, Philipps-University, Fb. 15-Chemie, Hans-Meerwein-Straße, D-35032 Marburg, Germany

AUTHOR EMAIL ADDRESS

szirah@mnhn.fr, carlos.afonso@upmc.fr, linne@chemie.uni-marburg.de, thomas.knappe@staff.uni-marburg.de, marahiel@chemie.uni-marburg.de, rebuffat@mnhn.fr, jean-claude.tabet@upmc.fr

RECEIVED DATE

TITLE RUNNING HEAD

CORRESPONDING AUTHOR FOOTNOTE

Séverine ZIRAH

Molécules de communication et adaptation des micro-organismes

FRE 3206 CNRS / Muséum National d'Histoire Naturelle

CP 54, 57 rue Cuvier

75005 Paris

FRANCE

Phone: +33 (0) 1 40 79 31 40

Fax : +33 (0) 1 40 79 31 35

Email: szirah@mnhn.fr

ABSTRACT

Lasso peptides constitute a class of bioactive peptides sharing a knotted structure where the *C*-terminal tail of the peptide is threaded through and trapped within an *N*-terminal macrolactam ring. The structural characterization of lasso structures and differentiation from their unthreaded topoisomers is not trivial and generally requires the use of complementary biochemical and spectroscopic methods. Here we investigated two antimicrobial peptides belonging to the class II lasso peptide family and their corresponding unthreaded topoisomers: microcin J25, which displays a typical lasso CID fragmentation pattern and capistruin, for which CID fragmentation does not permit to unambiguously assign the lasso structure. The two pairs of topoisomers were analyzed by ESI-FT-ICR-MS upon CID, IRMPD and ECD. Both CID and ECD spectra clearly permitted to differentiate MccJ25 from its non-lasso topoisomer, while for capistruin, only ECD was informative and showed different extent of hydrogen abstraction for the threaded and unthreaded topoisomers. This work shows the potentiality of ECD for structural characterization of peptide topoisomers, as well as the effect of conformation on hydrogen abstraction subsequent to ECD.

INTRODUCTION

Molecular knots constitute original structural motifs in biomolecules and fascinating topological objects.[1-2] Bioactive peptides with knotted structures, such as lasso peptides and cyclotides, share a knot topologically defined by covalent crosslinks (lactam ring and/or cysteine bridges) and display very compact and stable structures. Such structures therefore constitute valuable scaffolds with potential applications in peptide-based drug design.[3-4] Lasso peptides form a class of bacteria-produced bioactive peptides with a threaded structure involving an 8/9-residue ring resulting from an amide bond between the *N*-terminus and the carboxylic group of an Asp/Glu residue, crossed by the *C*-terminal tail.[5] They form three classes depending on the absence (class II) or presence of one (class III) or two (class I) disulfide bridges that further stabilize the structure.[6] They exhibit a variety of biological activities, involving enzyme inhibition, receptor antagonism, antimicrobial or anti-HIV activities (see Table 1 for class II lasso peptides).[5,7-12] Their compact structure confers to lasso peptides a great resistance to denaturing conditions and certain proteases. The structural characterization of lasso structures and differentiation from their unthreaded topoisomers is not trivial and generally requires the use of complementary biochemical and spectroscopic methods.

Microcin J25 (MccJ25) is a lasso peptide secreted by *Escherichia coli* AY25 that exerts potent antibacterial activity against *Escherichia* and *Salmonella* species, through import in the target bacteria upon interaction with the iron-siderophore receptor FhuA[13] and inhibition of the RNA polymerase.[14] Its *C*-terminal (18-21 SFYG) segment is sterically entrapped into the macrolactam ring by F19 and Y20 located on each side of the ring. Cleavage within the Y9-I18 region, either in solution by enzymatic cleavage or acidic hydrolysis, or in the gas-phase upon collision-induced dissociation (CID), generates two peptide complexes associated through the steric hindrance provided by the side chains of F19 and Y20.[15-17] A particular gas-phase fragmentation pattern upon CID, as compared to the corresponding synthetic peptide encompassing a macrolactam ring, has also been reported for the lasso peptide RES 701-1,[18] but generally only NMR could unambiguously characterize the threading of the *C*-terminal tail into the macrolactam rings.[7-8,19] Capistruin is a lasso peptide secreted by *Burkholderia thailandensis* E264 discovered by genome mining, which exhibits antibacterial activities against closely related *Burkholderia* and *Pseudomonas* species.[8] It showed a low overall fragmentation pattern upon CID. However, no two-peptide complexes unequivocally proving a lasso structure were observed.[8] Comparison of the three-dimensional structures of MccJ25 and capistruin shows that MccJ25 displays a large loop involving a short antiparallel β-sheet and a very short *C*-terminal segment below the ring, while capistruin displays a very tight loop and a much longer *C*-terminal segment below the ring (Figure 1). In both peptides, the steric entrapping of the *C*-terminal tail in the macrolactam ring relies on the presence of a bulky amino acid located below the ring: Y20 for MccJ25[20] and R15 for capistruin.[21] In addition, both peptides contain a bulky amino acid located above the ring: F19 for MccJ25 and R11 for capistruin. However, in capistruin the two bulky amino acids, R11 and R15, are not adjacent.

Fourier transform ion cyclotron resonance (FT-ICR) mass spectrometry has shown a great capability for the analysis of peptides and proteins, in particular for the characterization of post-translational modifications. The main advantages of this method are the high resolution provided by the ICR cell together with the multiple excitation modes available on this instrument, such as (i) collision-induced dissociation (CID) in the external collision cell or in the ICR cell through sustained off resonance irradiation (SORI) under CID conditions,[22] (ii) infrared multiphoton dissociation (IRMPD)[23] and (iii) for positive ions, electron-capture dissociation (ECD).[24-25] The latter has revealed particularly appealing for the structural characterization of peptides and proteins[26], given the ability to cleave the peptide backbone while leaving intact labile side-chain modifications, or without disrupting the non-covalent interactions involved in the three-dimensional structure of proteins.[27-28] In order to develop an analytical method to unambiguously characterize topoisomeric peptides by mass spectrometry, the fragmentation patterns MccJ25, capistruin, and their corresponding non-lasso topoisomers MccJ25-lcm and capistruin-lcm, were analyzed by ESI-FT-ICR-MS upon CID, SORI-CID, IRMPD and ECD.

EXPERIMENTAL SECTION

*Peptides*

MccJ25 was produced from a culture of *E. coli* MC4100 harboring the plasmid pTUC202, cultivated for 16 h in M63 medium supplemented with 1 mg/mL vitamin $B_1$, 0,02% $MgSO_4$, 0,02% glucose and 1 g/L casamino acids. MccJ25 was purified from the culture supernatant by solid phase extraction on a SepPak C18 35 cc cartridge and by semi-preparative RP-HPLC on a C18 µBondapak column (300 mm × 3.9 mm, 10 µm, Waters).

Capistruin was produced from a culture of *Burkholderia thailandensis* E264 incubated for 24 h at 42 °C in M20 medium containing gentamycin (8 µg/mL), as described previously.[8] The purification was carried out by solid phase extraction followed by preparative RP-HPLC.[8]

The synthetic cyclic branched peptides, MccJ25-lcm and capistruin-lcm, were obtained from Genepep (Montpellier, France).

*FT-ICR mass spectrometry*

Mass spectrometry experiments were performed on a hybrid Qh-FT/ICR (Bruker Daltonics, Bremen, Germany) equipped with an off-axis Apollo II ESI source operated in positive ion mode. A syringe pump was used to infuse the peptide solution (10 µM in $H_2O/CH_3CN$ + 0.1% (v/v) formic acid) with a gas-tight syringe at a flow rate of 120 µL/h. CID experiments were performed in the collision cell or upon SORI-CID in the ICR cell using argon as collision gas. For SORI-CID, the ions were activated using a -1000 Hz frequency offset for 250 ms with a 3.3 Vp-p excitation amplitude. A pumping delay of 2 s was applied prior to ion analysis. When necessary, $MS^3$ experiments were conducted using CID in the collision cell in conjunction with either SORI-CID or IRMPD. ECD experiments were performed with an indirectly heated hollow cathode set to 1.7 A. Electrons emitted during 0.1 s were injected into the ICR cell with a 1.2 V bias.

*Nomenclature of the product ions*

To avoid any ambiguity, the nomenclature introduced by Roepstorff and Folman,[29] further modified by Biemann and coworkers,[30] was adopted to describe the product ions. For internal fragment ions, the general presentation form is $(b_r y_s)_{(r+s-t)}$, where r, s and t indicate the number of amino acids cleaved counting from the *N*-terminus, the number of amino acids cleaved from the *C*-terminus and the total number of amino acid residues in the peptide, respectively. The subscript (r+s−t) thus accounts for the number of residues in the internal fragment. As an example, the internal fragment GIGTPI of MccJ25 has r = 17, s = 10 and t = 21 and is denoted $(b_{17}y_{10})_6$. The product ions resulting from fragmentation in the loop of the lasso structure and where the *C*-terminal tail is threaded into the macrolactam ring were denoted $[(b_r)*(y_s)]$, where r and s indicate the number of amino acids in the non-covalently bound *N*-terminal and *C*-terminal peptide segments. As an example, the product ion [MccJ25-{GIGTPI}] is denoted $[(b_{11})*(y_4)]$. For ECD fragmentations, similar notation using c/z codes were used. The charge states were omitted for singly-charged species and indicated for doubly- and triply-charged species.

RESULTS AND DISCUSSION

*Fragmentation patterns of MccJ25 and capistruin and their non-lasso topoisomers generated by CID, SORI-CID and IRMPD*

The CID spectra of MccJ25, capistruin, and their corresponding synthetic non-lasso topoisomers, MccJ25-lcm and capistruin-lcm, which contain the macrolactam ring but without steric entrapping of the *C*-terminal tail inside, were obtained through CID in the external collision cell or SORI-CID in the ICR cell. The spectra provided similar information, therefore only SORI-CID spectra are presented, since these experiments permit to avoid *m/z* discrimination during the ion transfer between the collision cell and the ICR cell.

The SORI-CID spectrum of the $[M+3H]^{3+}$ species of MccJ25 (*m/z* 703.0) showed a complicated fragmentation pattern, involving two competitive dissociation processes (Figure 2A, supplementary Table S1). On the one hand, extensive fragmentations in the Y9-I17 loop were detected, yielding typical two-peptide product ions resulting from the steric entrapping of the (SFYG) *C*-terminal segment in the macrolactam ring, as already reported.[15,17] The main product ions consisted of $[(b_{16})*(y_4)]^{3+,2+}$ and $[(b_{15})*(y_4)]^{3+,2+}$, with consecutive $H_2O$ and $CO$ neutral losses, together with $[(b_{11})*(y_6)]^{2+}$ and $[(b_{11})*(y_4)]^{2+}$. Several internal product ions, complementary to the two-peptide product ions, were also detected: $(b_{15}y_{10})_4$, $(b_{17}y_{10})_6$, $(b_{19}y_8)_6$, $(b_{16}y_9)_4$... On the second hand, complementary y/b series corresponding to fragmentations in the C-terminal tail were detected. They consisted of the $b_8$, $b_9$, $b_{10}$, $b_{11}^{1+, 2+}$, $b_{12}$, $b_{13}^{1+, 2+}$, $b_{19}^{2+}$, $b_{20}^{3+}$ series, together with the $y_2$, $y_3$, $y_4$ and $y_6$ series (supplementary Table S1). These series indicated a release of the (SFYG) C-terminal segment and therefore that fragmentation in the macrolactam ring occurred. Alternatively, such product ions could result from consecutive dissociation from $b_{20}^{3+}$. Indeed, the loss of the C-terminal G21 residue may release the C-terminal tail. Finally, consecutive $H_2O$ and $CO$ losses from the precursor ion were observed ($[M+3H-H_2O]^{3+}$ being the base peak of the spectrum). The SORI-CID spectrum of the $[M+2H]^{2+}$ species of MccJ25 displayed a weak dissociation profile, with $[M+2H-H_2O]^{2+}$ as base peak (data not shown). The $[M+3H]^{3+}$ species of MccJ25-lcm showed a more specific fragmentation pattern under SORI-CID than that of its lasso topoisomer (Figure 2B, supplementary Table S2). As for MccJ25, the base peak was $[M+3H-H_2O]^{3+}$ at m/z 697.0. The spectrum was dominated by the b series (singly-charged $b_9$ to $b_{13}$, doubly-charged $b_{11}^{2+}$ to $b_{15}^{2+}$, $b_{17}^{2+}$, $b_{19}^{2+}$, triply-charged $b_{20}^{3+}$). Intense product ions corresponding to one or two $H_2O$ losses were observed for $b_{17}^{2+}$, $b_{19}^{2+}$ and $b_{20}^{3+}$. This behavior is most likely due to the presence of the T15 and S18 residues in the vicinity of the charge. The complementary $y_2$, $y_4$, $y_6$, $y_7$ and $y_8$-$H_2O$ were also observed. Finally, only few internal product ions were detected, such as $(b_{19}y_8)_6$-$2H_2O$ (m/z 567.3). As expected, two-peptide product ions were not detected.

The IRMPD spectrum of the $[M+3H]^{3+}$ species of MccJ25 (m/z 703.0) revealed to be less complex than the corresponding SORI-CID spectrum (Figure 3A, supplementary Table S3). The spectrum was dominated by the b series ($b_8$ to $b_{13}$, $b_{16}^{2+}$, $b_{17}^{2+}$ and $b_{19}^{2+}$ with one or two $H_2O$ losses). The y series consisted in $y_2$, $y_3$, $y_4$, $y_6$. Only few low abundance two-peptide product ions were detected: $[(b_{16})*(y_4)-H_2O]^{3+,2+}$ (m/z 659.3, 988.5) and $[(b_{15})*(y_4)-H_2O]^{2+}$ (m/z 939.9). This result contrasts with the SORI-CID spectrum where these two peptide ions are produced in high abundance. In addition, several internal product ions were observed, in particular $(b_{10}y_{13})_2$ and/or $(b_{20}y_3)_2$ (*m/z* 311.1), together with $(b_{17}y_{10})_6$-$H_2O$ (*m/z* 521.3). This fragmentation pattern illustrates that IRMPD triggers consecutive dissociation that result in multiple neutral losses and in the opening of the macrolactam ring, and therefore in the release of the entrapped *C*-terminal segment. A similar behavior towards IRMPD was observed for the $[M+2H]^{2+}$ species of MccJ25 (data not shown). The IRMPD spectrum of the $[M+3H]^{3+}$ species of MccJ25-lcm showed a y/b fragmentation pattern close to that of MccJ25, but with a significantly higher fragmentation efficiency (Figure 3B, supplementary Table S4). While for the lasso peptide MccJ25 the precursor ion dominated the IRMPD spectrum, for the non-lasso topoisomer, the $b_{13}$ product ion was the most intense ion. The IRMPD spectra of MccJ25 and MccJ25-lcm were dominated by b/y series corresponding to cleavages within the *C*-terminal tail but were fairly different in terms of product ion relative abundance. The $b_{17}$ to $b_{20}$ product ions, corresponding to cleavages within the (SFYG) *C*-terminal segment, were favored for the lactame topoisomer as compared to the lasso. This can be explained by the burring of this segment in the macrolactame ring, which may strengthen the peptidic bonds in this region. Only the minor two-peptide product ions characteristic of the lasso topoisomer, *i.e.*, $[(b_{16})*(y_4)-H_2O]^{3+,2+}$ and $[(b_{15})*(y_4)-H_2O]^{2+}$, could allow to unambiguously distinguish the two structures.

SORI-CID is an "almost" resonant activation mode where only the precursor ion is effectively activated at its cyclotron frequency. With IRMPD, both the precursor ion and product ions present in the laser beam path are activated, which tend to enhance consecutive dissociation.[23] Thus, the $[(b_r)*(y_s)]$ two-peptide product ions can dissociate, yielding the conventional $y_s$ or $b_r$ product ions. Consecutive dissociation can also explain the high abundance of some internal product ions such as $(b_{19}y_8)_6$-$2H_2O$ (m/z 567.3) (Figure 2B, 3B). In this regard, the SORI-CID spectrum presents more information about the existence of the lasso form, given the presence of the two-peptide product ions and although in this case the dissociation efficiency is lower as loss of water dominates the product ion spectrum. Thus, resonant activation mode that limits consecutive dissociation is more adequate for the structural characterization of lasso peptides as it allows here to conserve intact two-peptide ions that are a direct proof of the threaded structure.

The SORI-CID spectra of capistruin and capistruin-lcm (supplementary Figure S1) were poorly informative, as they showed mainly consecutive $H_2O$ and $NH_3$ losses from the precursor ion. In addition, capistruin showed very weak y/b product ions resulting from cleavages in the *C*-terminal tail. The IRMPD spectra of capistruin and capistruin-lcm (Figure 4) also revealed mainly neutral losses, together with weak y/b series corresponding to cleavages in the *C*-terminal tail that were also affected by extensive neutral losses, and internal product ions. Many common product ions were detected for both topoisomers, but with different relative abundance and neutral loss patterns. As an example, the $[b_{15}-H_2O]^{2+}$, $[b_{16}-H_2O]^{2+}$, $[b_{17}-H_2O]^{2+}$, $[b_{18}-H_2O]^{2+}$ series dominated the *m/z* 700-1000 region in the IRMPD spectrum of capistruin, while this region was dominated by $y_6$-$NH_3$, $y_7$-$NH_3$, $y_8$-$NH_3$ for capistruin-lcm. This trend suggests that the product ions $[b_{15}-H_2O]^{2+}$ to $[b_{18}-H_2O]^{2+}$, more intense for the lasso topoisomer, correspond to lasso-structured product ions, maintained through R11 and R15 side-chain located above and below the ring. Thus, as for McCJ25, the ergodic gas-phase fragmentations of capistruin appear to result in two dissociative processes, (i) the former preserving the lasso structure and yielding the $[b_{15}-H_2O]^{2+}$ to $[b_{18}-H_2O]^{2+}$ product ions, and (ii) the latter yielding y/b series corresponding to cleavages in the *C*-terminal tail, which indicates a previous precursor ion isomerization through opening of the macrolactam ring. However, the short sequence of the loop did not permit to generate two-peptide product ions, which require two bond cleavages in the loop region. Contrary to McCJ25 and its non-lasso topoisomer, the peptides capistruin and capistruin-lcm were not distinguished by CID, which was completely dominated by the water loss, but gave different product ion abundance profiles upon IRMPD, which suggests the formation of two-peptide product ions for the lasso topoisomer. This weak fragmentation behavior upon ergodic dissociative processes is attributed to the presence of R11 and R15 residues in the tail region, which most probably sequester the protons. The lasso peptides McCJ25 and capistruin displayed a very extensive or very weak fragmentation extent, respectively, which illustrates the influence basic amino acids on ergodic dissociation, in agreement with the mobile proton model.[31]

*Fragmentation patterns of MccJ25 and capistruin and their non-lasso topoisomers generated by ECD*

The ECD spectra of MccJ25 revealed different fragmentations patterns for the $[M+2H]^{2+}$ and $[M+3H]^{3+}$ species (Figure 5 A,B). The doubly-charged species generated mainly a c series ($c_{17}$, $c_{18}$, $c_{19}$, $c_{20}$), together with two-peptide product ions: $[(c_9)*(y_7)-H_2O]$, $[(c_{10})*(y_7)-H_2O]$ and $[(c_{11})*(y_7-H_2O)]$. Two relatively intense product ions at *m/z* 1298.7 and 1341.7 could not be assigned. The triply-charged species yielded a y series ($y_4$, $y_6$, $y_8$) and b series ($b_{10}$, $b_{11}$, $b_{13}$, $b_{15}$) corresponding to + 1 u mass shifts, noted $b"_n$ (Figure 5B). Almost no c/z• ions were detected, with the exception of $c_{18}$ (*m/z* 1738.8) and $c_{19}^{2+}$ (*m/z* 943.5). In addition, in a lower extent, two-peptide product ions such as $[(b_8)*(y_{12})]^{2+}$ and the series $[(c_8)*(y_4)]$, $[(c_9)*(y_4)]$ and $[(c_{10})*(y_4)]$ were present. Finally, c/y or b/y internal product ions corresponding to tail fragments, *i.e.*, $(c_{19}y_4)_2$, $(c_{20}y_4)_3$, $(c_{19}y_6)_4$, $(c_{19}y_8)_6$, …, were detected. The non-lasso topoisomer MccJ25-lcm, dissociated upon ECD, showed certain trends common to its lasso topoisomer: a $c_{11}$ to $c_{20}$ series for the doubly-charged species (Figure 5C), and two competitive dissociation processes for the triply-charged species (Figure 5D): (i) ECD in the ring and consecutive y/ b"• series and (ii) ECD in the tail yielding a c/z• series. As expected, the two-peptide product ions, characteristic of the lasso structure, were absent. Note that the dissociation efficiency of the charge-reduced species was significantly higher for the non-lasso peptide as compared to the lasso topoisomer. This illustrates the high stability of the lasso structure, as observed by CID.

The almost complete absence of c/z• ions for the $[M+3H]^{3+}$ species of MccJ25 and MccJ25-lcm is quite unusual and should indicate a particular dissociation pathway. Indeed, electron capture is expected to yield charge-reduced species $[M+3H]^{•2+}$ that consecutively undergo c/z• cleavage. The b"• series and bipeptic product ions permitted to propose a competitive fragmentation pattern involving two isomeric charge-reduced species $[M+3H]^{•2+}$ (*m/z* 1054.5) (Scheme 1). Most likely, the initial ECD occurred within the macrolactam ring, which is consistent with the presence of the basic histidine residue at position 5 (pathway a). The excess of energy in the $[M+3H]^{•2+}$ charge-reduced species, due to electron capture, would yield consecutive dissociation involving a conventional y/b cleavage within the *C*-terminal tail (b series corresponding to + 1 u mass shifts). On the other hand, an initial ECD in the *C*-terminal tail (pathway b) would result in a two-peptide charge-reduced species [M+3H]$^{•2+}$, further dissociated into two-peptide and internal product ions. Such pathways result from the presence of both a cyclic and a linear region in the peptide. Multiple backbone cleavages in cyclic peptides have been proposed to result from a free radical reaction cascade in which the α-carbon radical formed by electron capture propagates along the peptide backbone by free radical rearrangements.[32] For McJ25 and MccJ25-lcm, which are both cyclic-branched peptides, the dissociation appear more complex, with both radical-driven and charge-driven bond cleavages. Such a competition between these two processes has been described recently for arginine-containing peptides, and the absence of basic residues is proposed to favor charge-driven fragmentations while arginine would promote radical-driven fragmentations.[33] The absence of basic residues in the tail of MccJ25 supports this tendency.

For both CID and ECD, MccJ25 showed a fragmentation pattern typical of the lasso structure, which permits to assign the lasso character directly, even without a comparison to the non-lasso topoisomer. The formation of the typical two-peptide product ions is particularly favorable for MccJ25, as compared to capistruin given the large size of the loop above the ring and the vicinity of the two bulky amino acids that maintain the *C*-terminal tail entrapped within the ring. Many two-peptide product ions show $y_4$ (SFYG) as minimal entrapped segment.

The dissociation of the [M+3H]$^{3+}$ and [M+2H]$^{2+}$ species of capistruin and its topoisomer capistruin-lcm were also investigated under ECD conditions. The ECD spectra of the [M+2H]$^{2+}$ species of capistruin and capistruin-lcm (supplementary Figure S2) revealed similar fragmentation patterns, with a $c_{16}$ to $c_{18}$ series. The ECD spectrum of the [M+3H]$^{3+}$ species of capistruin (Figure 6A) showed an abundant charge-reduced species [M+3H]$^{•2+}$, together with product ions resulting from neutral losses and a series of c and z$^•$ ions. The product ion at *m/z* 996 corresponded to a loss of $CH_5N_3$ from the arginine side chain, a usual side chain cleavage in ECD.[34] The $z_5^•$ to $z_9^•$ product ions corresponding to cleavages within the tail were fairly abundant. The complementary $c_{11}$ to $c_{18}$ were produced as singly-charged, but also as doubly-charged species for the larger product ions ($c_{15}$ to $c_{18}$). $c_{11}$, the smallest c ion detected, corresponded to the cleavage after R11. The $z_{10}^{\bullet}$ and $z_9^{\bullet}$ product ions were detected, but in very low abundance. It should be noted that none of these main cleavages involves dissociation of the ring. This indicates that an initial cleavage produced after electron capture occurred within the tail. This is most likely due to the presence of the two basic residues (R11 and R15) in this region, which promotes local c/z$^{\bullet}$ fragmentations in ECD. However, some low abundance product ions were attributed to internal cleavages within the ring. The ECD spectrum of the $[M+3H]^{3+}$ species of the capistruin-lcm topoisomer revealed a similar fragmentation pattern (Figure 6B), but with significant differences in the relative abundance of the product ions. In particular, the $z_{10}^{\bullet}$ and $z_9^{\bullet}$ product ions were detected in a much larger abundance as compared to capistruin. This trend suggests that when these cleavages occur for the lasso topoisomer capistruin, the two peptidic entities constitute two-peptide product ions as for MccJ25, due to the steric hindrance provided by R11 and R15 (as already suggested for CID experiments).[8] In spite of this specific profile of product-ion abundances, no strong evidence of the lasso structure could be detected.

Interestingly, part of the product ions corresponded to c and z$^{\bullet}$, but several c$^{\bullet}$ and z were also detected. Such a trend is relatively frequent during ECD dissociation of peptides and it has been proposed that the c$^{\bullet}$ and z species are produced from a hydrogen transfer occurring in a long life c/z$^{\bullet}$ complex.[35-37] For capistruin and capistruin-lcm topoisomers, the c$^{\bullet}$ ions revealed particularly abundant for cleavages of the tail nearby the ring. In addition, the relative abundance of c$^{\bullet}$/z species was strongly reduced for the non-lasso topoisomer. This behavior is illustrated in the spectra enlargement in Figure 6C. In the non-lasso peptide, the normal $c_{11}$ was mainly detected, while a significant amount of $c_{11}^{\bullet}$ species was detected for the lasso topoisomer. The same behavior was observed for $z_9$, which was mostly detected as $z_9^{\bullet}$ for the capistruin-lcm and as a mixture $z_9^{\bullet}/z_9$ for capistruin. Only a comparison with the non-lasso topoisomer permitted to assign the lasso structure of capistruin, with regard to the relative abundance of the product ions and the extent of hydrogen abstraction. Therefore, ECD appeared as more powerful in differentiating the two topoisomers capistruin and capistruin-lcm as compared to ergodic dissociation and generated mainly c/z• product ions within the *C*-terminal tail. The absence of y/b product ions suggested that electron capture occurred within the tail and yielded a free radical reaction cascade resulting in multiple backbone cleavages.

CONCLUSIONS

Our data showed that characterization of lasso peptides from gas-phase dissociation can be more or less complex depending on several parameters (Figure 1): (1) the size of the loop above the macrolactam ring, (2) the vicinity of the bulky amino acids in the *C*-terminal tail located above and below the macrolactame ring, (3) the presence of basic residues within the ring and the tail (that limits the fragmentation extent in ergodic dissociation and promote local fragmentations in ECD). ECD appears particularly promising, since it provides evidence of the lasso structure for both McJ25 and capistruin. It generates competition between charge-driven and radical-driven bond cleavages, which orientation was strongly influenced by the presence or absence of arginines. The presence of two-peptide product ions is the main trend typical of the lasso structure and was clearly observed upon CID and ECD of McJ25. For capistruin, the absence of double cleavage between D9 and R11 segment did not permit to detect it.

The different hydrogen abstraction extent for the two topoisomers capistruin and capistruin-lcm is the first evidence of the influence of topology on this process, which was mainly interpreted with regard to the amino acid sequences. It suggests potential applications in the characterization of peptide or protein conformations, topology or non-covalent associations.

ACKNOWLEDGEMENT

This work was supported by the ANR project no. BLAN_NT09_692063 and the Deutsche Forschungsgemeinschaft (DFG MA 811-25/1).

# FIGURES AND TABLES

## Table 1

Sequences and biological activities of type-II lasso peptides[a]

| Name | Sequence | Producing bacteria | Biological activity | Reference |
|---|---|---|---|---|
| **RES-701-1** | 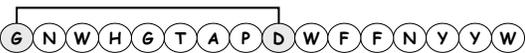 | *Streptomyces* sp. RE-701 | Endothelin type B receptor selective antagonist | 9 |
| **Anantin**[b] | 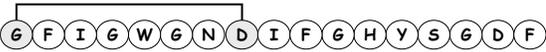 | *Streptomyces coerulescens* | Atrial natriuretic factor antagonist | 12 |
| **Propeptin**[b] | 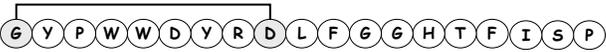 | *Microbispora* sp. SNA-115 | Prolyl endopeptidase inhibitor | 11 |
| **Lariatins** | 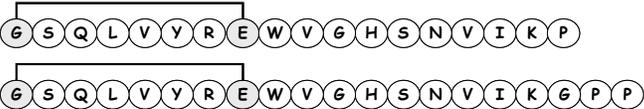 | *Rhodococcus* sp. K01-B0171 | Antimycobacterial | 7 |
| **Microcin J25** | 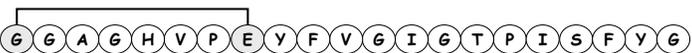 | *Escherichia coli* AY25 | Antibacterial RNA polymerase inhibitor | 10,15,17 |
| **Capistruin** | 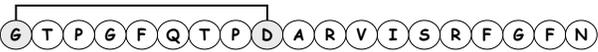 | *Bhurkholderia thailandensis* E264 | Antibacterial | 8 |

[a] The sequences are represented unthreaded for clarity.

[b] The lasso structure (threading of the tail into the macrolactame ring) is hypothesized but has not been shown.

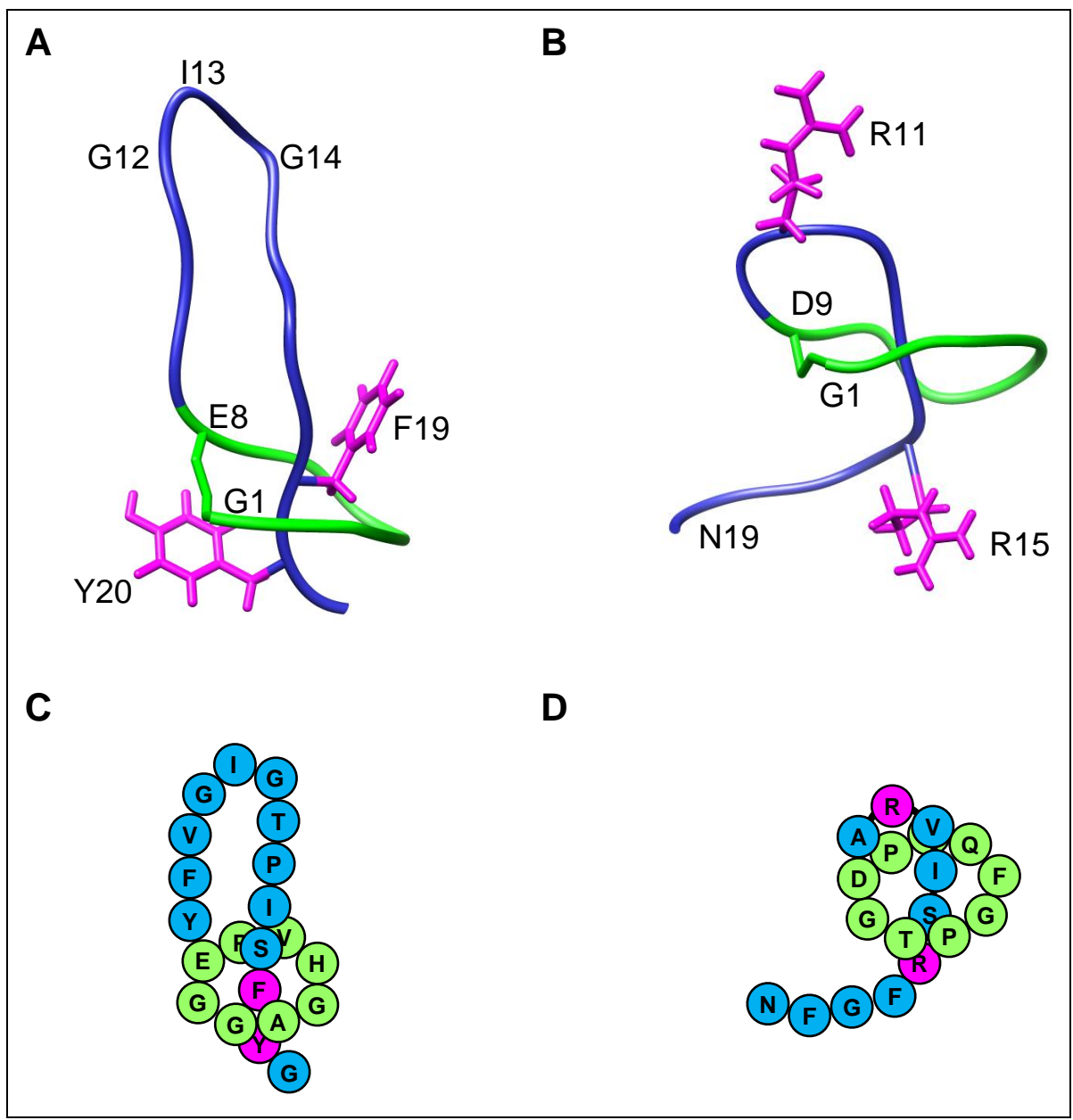

**Figure 1.** Three-dimensional structures of MccJ25[17] (A) and capistruin[8] (B). Scheme representing MccJ25 (C) and capistruin (D), showing the macrolactam ring in green, the *C*-terminal tail in blue and the bulky residues located above and below the macrolactame ring in magenta.

**Figure 2.** SORI-CID spectra of the [M+3H]$^{3+}$ species of (A) MccJ25 and (B) MccJ25-lcm (*m/z* 703.0). Typical product ions are displayed on the right (the neutral losses are not indicated on the schemes for clarity).

**Figure 3.** IRMPD spectra of the $[M+3H]^{3+}$ species of (A) MccJ25 and (B) MccJ25-lcm (*m/z* 703.0). The main product ions are displayed on the right (the neutral losses are not indicated on the schemes for clarity).

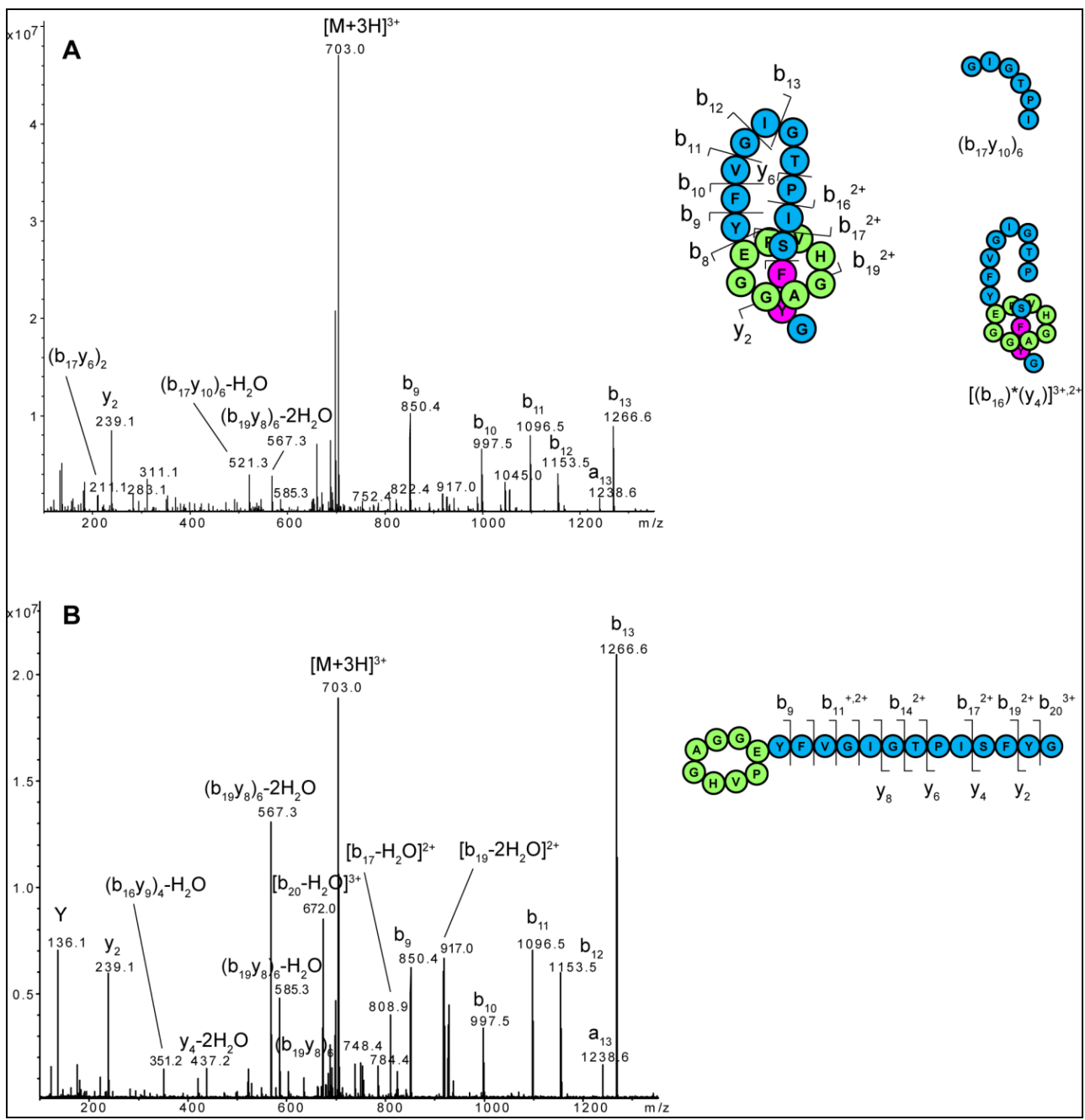

**Figure 4.** IRMPD spectra of the [M+3H]$^{3+}$ species of (A) capistruin and (B) capistruin-lcm (*m/z* 683.7). The main product ions are displayed on the right (the neutral losses are not indicated on the schemes for clarity).

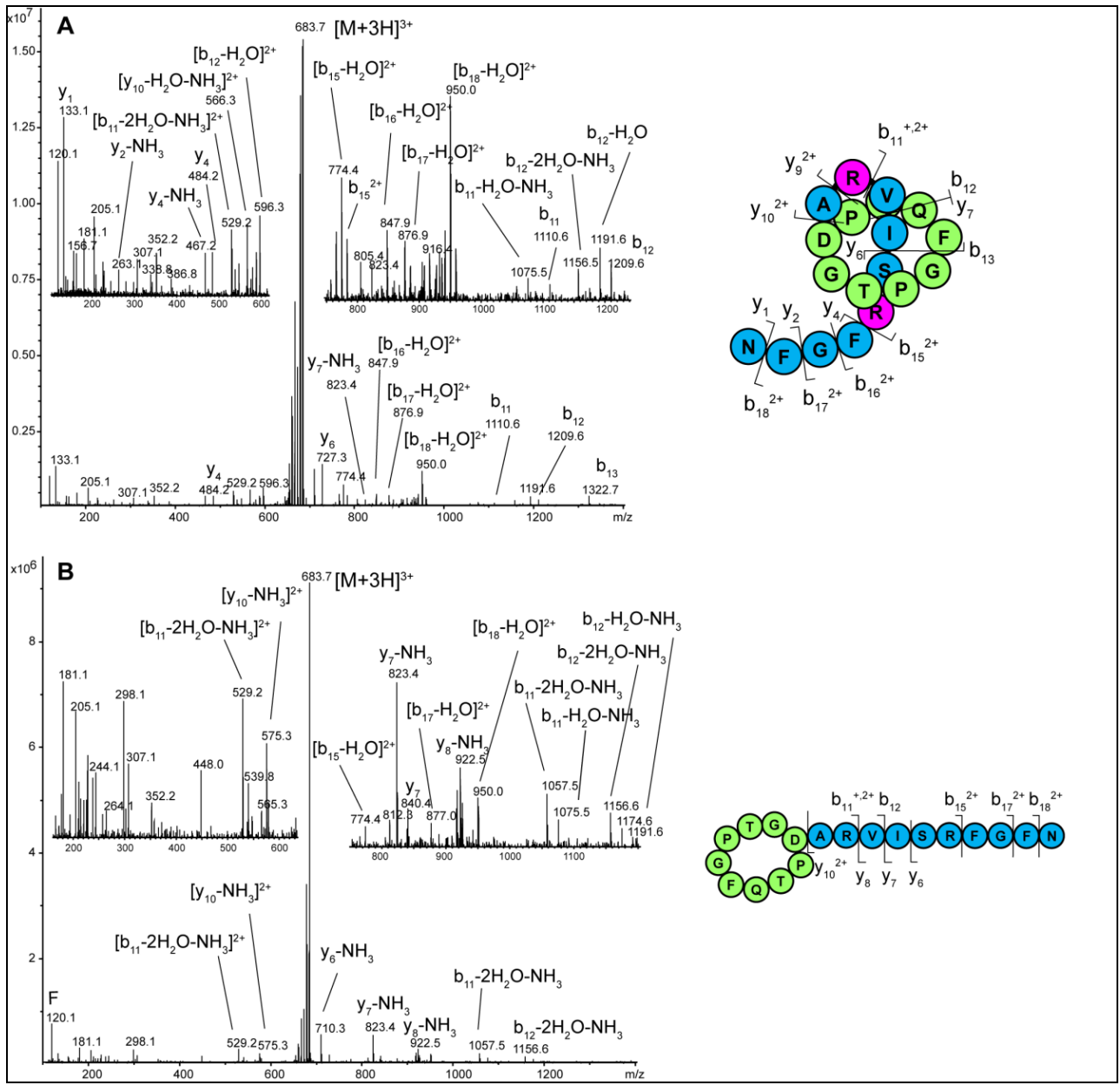

**Figure 5.** ECD spectra of (A) [M+2H]$^{2+}$ of MccJ25 (*m/z* 1054.0) (B) [M+3H]$^{3+}$ of MccJ25 (*m/z* 703.0), (C) [M+2H]$^{2+}$ of MccJ25-lcm (*m/z* 1054.0) and (D) [M+3H]$^{3+}$ of MccJ25-lcm (*m/z* 703.0). The main product ions are displayed (the neutral losses are not indicated on the schemes for clarity).

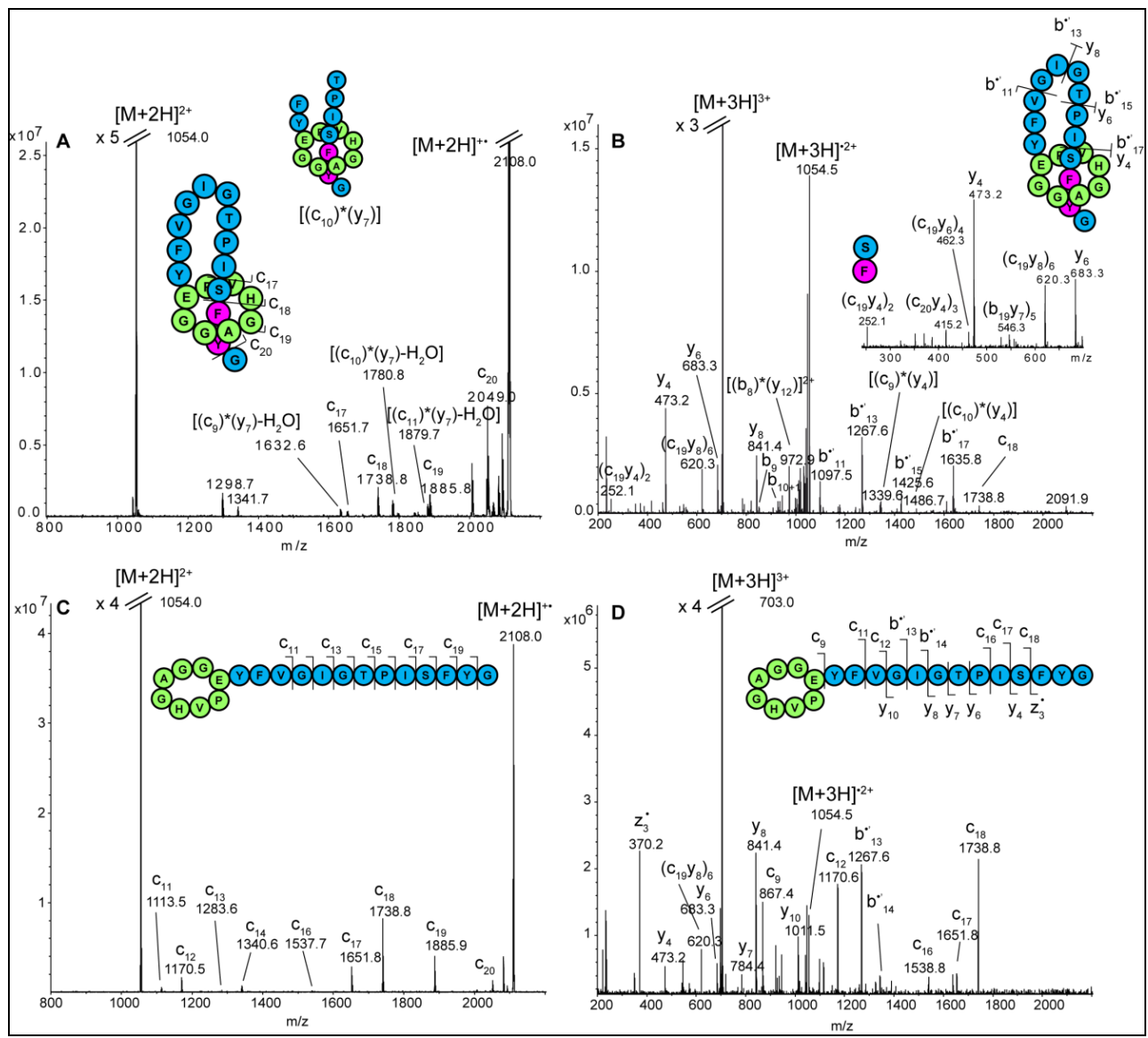

**Figure 6.** ECD spectra of the $[M+3H]^{3+}$ species of (A) capistruin (B) and capistruin-lcm (*m/z* 683.7). Typical product ions are displayed. (C) $c_{11}/c_{11}^{\bullet}$ isotopic profiles obtained experimentally for capistruin and capistruin-lcm (on the left), and expected for $c_{11}$ and $c_{11}^{\bullet}$ (on the right).

**Scheme 1.** Pathways proposed for the ECD of the [M+3H]$^{3+}$ species of MccJ25.

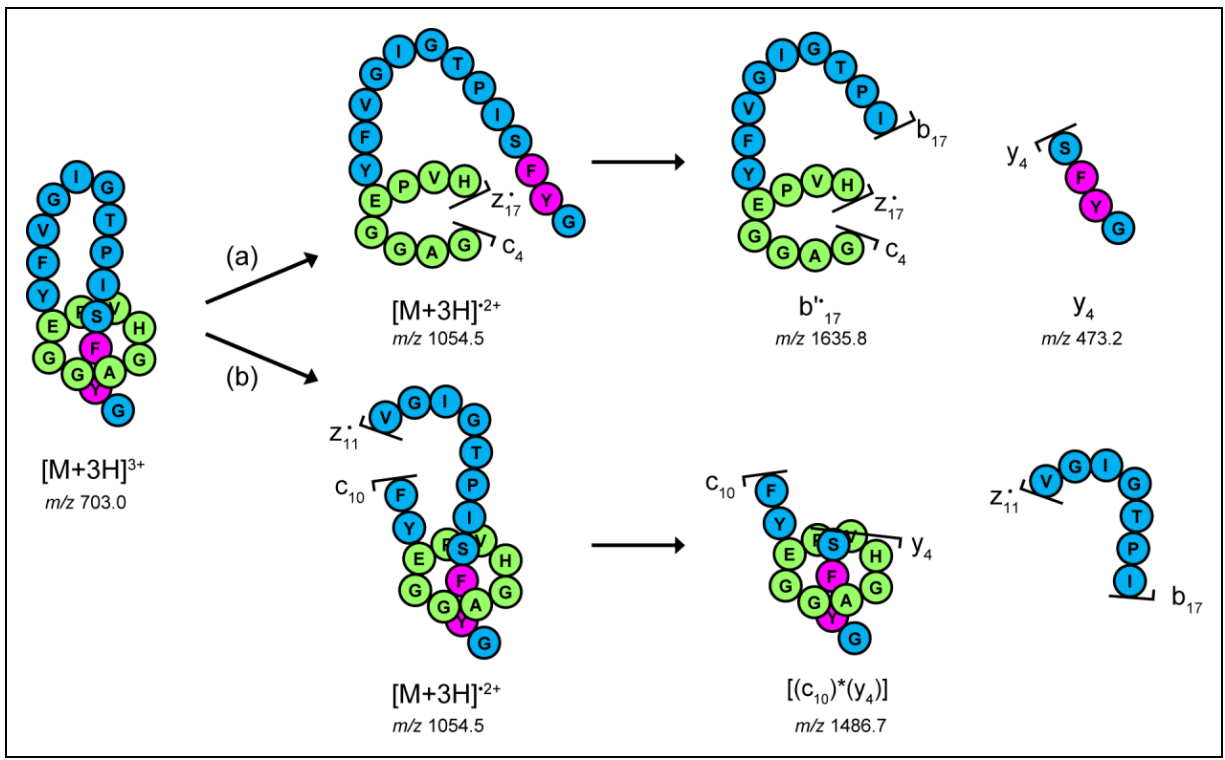